# Surface-related white light emission phenomenon in transparent solids

M. Chaika[*], R. Tomala, M. Oleszko, W. Strek

*Institute of Low Temperature and Structure Research Polish Academy of Science, Okolna 2, 50-422 Wroclaw, Poland.*

m.chaika@intibs.pl

**Abstract:** Laser induced white emission (LIWE) caused by infrared laser excitation in Cr:YAG transparent ceramics was investigated. It was found that ceramics generates bright LIWE for excitation powers above a critical threshold. The LIWE was observed on the surface but not in the bulk on both sides of the sample. The vacuum conditions are required to observe LIWE. This phenomenon was discussed within the frame of Inter-Valence Charge Transfer (IVCT) mechanism in the $Cr^{3+}/Cr^{4+}$ ion pair.

## Introduction

Anti-Stokes photoluminescence excited by near-infrared lasers has been widely observed in various optically active materials including ceramics, single crystals, nanopowders, organic dye solutions, etc [1–3]. An example of the anti-Stokes photoluminescence is the Laser Induced White Emission (LIWE) which has been observed for samples placed in vacuum and excited by focused near-infrared laser beam. Recent LIWE studies have attracted a considerable interest. This phenomenon was observed in different materials including inorganic materials, hybrid nanostructures, organometallics, etc. Generally, LIWE can be detected when the density of laser beam exceeds $10^2$ W/cm$^2$. LIWE covers whole Vis and NIR part of the spectrum and its intensity can be regulated both by the excitation power and ambient pressure. Spectral shape of LIWE differ from the shapes of luminescent bands obtained during normal upconversion process, in which the emission bands can be ascribed to discrete transitions of activators and are characterized by continuum spectra extending to the infrared region [4].

The main feature of this phenomenon is the similarity of the spectra for different materials including non-luminescent ones like carbons, silica powders and other [1]. So, the white light emission can be treated as a general photo-physical process. The most popular models for described LIWE include black body radiation, photon avalanche, electron-hole recombination, RE-$O^{2-}$ charge transfer, Inter Valence Charge Transfer (IVCT), and others [1–9]. Most of these theories are based on electron transfer in donor/acceptor pair. Due to the fact that the samples were excited by a focused laser beam, the thermal origin of white light emission is still possible, especially for nanopowder samples with low thermal conductivity. Therefore, at present there is no general model that could describe the optical properties of laser-driven white light emission. One of the reasons for this is the insufficient amount of experimental data on LIWE in transparent materials that severely limits the current understanding of this phenomenon.

The aim of this work is to expand the current knowledge on LIWE in order to understand the mechanism behind this phenomenon. So far, most studies were focused on opaque nanopowder materials. In contrast to these studies, here were discuss LIWE in transparent Cr:YAG ceramics.

## Materials and Methods

Cr:YAG ceramics were obtained from CoorsTek research laboratory located in Uden, Netherlands. LIWE was excited by a 975 nm laser diode in vacuum chamber. For emission spectra measurements AVS-USB2000 Avantes Spectrometer was used. The maximum laser output was 3 W with laser density at the surface of the sample reaching up to $1.5 \cdot 10^4$ W/cm$^2$.

## Results

The optical measurements have shown that Cr:YAG ceramics contain $Cr^{3+}$ and $Cr^{4+}$ ions. The linear transmission of the samples was 3.7 % and 80% at 1064 and 2000 nm, respectively. The sample thickness was 3.8 mm. The strong absorption in the visible and near-infrared is caused by $Cr^{4+}$ ions. The optical properties of Cr:YAG transparent ceramics are described in more detail in our previous work [10].

The Cr:YAG ceramics were placed into a quartz vacuum tube and irradiated by a focused Nd:YAG laser (**Fig. 1(a)**). The light emission spectrum of the sample is shown in **Fig. 1(b)**. LIWE consists of the band in visible and near infrared regions between 28000 cm$^{-1}$ (350 nm) and 10000 cm$^{-1}$ (1000 nm). LIWE was recorded from the laser irradiation spot on the sample surface [7]. No white light emission was detected from the bulk of the sample. During experiments, the laser beam enters and exits the sample volume from the other side. Unexpectedly, in some cases, LIWE was detected on the other side of the sample in the spot corresponding to laser beam exit.

The LIWE has shown a threshold behavior, as an increase in the laser power to 0.2 W/cm$^2$ did not lead to appearance of LIWE, and only when this value was exceeded, a weak signal was detected. Further increase in laser power leads to the strong rise in the emission intensity. Using the logarithmic dependence of LIWE intensity on the excitation energy, it was shown that at least 4 photons are involved into LIWE process which corresponds to ~ 38000 cm$^{-1}$. The LIWE phenomenon was observed only under low pressure conditions. The pressure dependence of the LIWE intensity exhibits threshold behavior, and no LIWE was observed at atmospheric pressure. Pressure decrease leads to increasing LIWE intensity. The decrease of the pressure from 0.1 to $5 \cdot 10^{-5}$ mbar does not affect the emission intensity and some kind of saturation is observed [6]. LIWE is reduced by two orders at lowering the ambient pressure below 0.1 mbar.

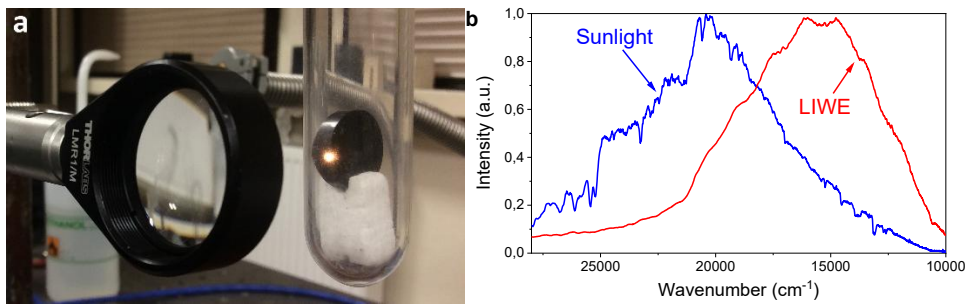

Fig. 1. Photo (a) and emission spectra (b) of Laser Induced White Emission (LIWE) from Cr:YAG transparent ceramics.

One of the features of LIWE in transparent materials is relatively short rise and decay times. The measured rise and decay times of LIWE in $Cr^{4+}$:YAG transparent ceramics were ~26 ms and ~17 ms, respectively, in agreement with the earlier reported data [6,11,12]. The data shown in **Fig. 2** represent the general trend, while in some cases the times of decrease/increase in the emission intensity were shorter than the time of measurement (~2 ms). Interestingly, materials with the high thermal conductivity ($Cr^{4+}$:YAG transparent ceramics [6], graphene ceramics [11], tungsten filament [13], etc. ) are characterized by two orders of magnitude shorter increase/decrease LIWE times compared to materials with low thermal conductivity (Yb:YAG nanopowders [8], Yb:$Sr_2CeO_4$ [14], $Yb^{3+}$-doped porous silica [2], etc.). Probably, such a difference is caused by negative influence of the temperature on LIWE process. Lower thermal conductivity causes faster accumulation of heat in the excitation spot allowing reaching higher temperature compared to the sample with higher thermal conductivity. For example, the temperature of the host during LIWE in transparent $Cr^{4+}$:YAG ceramics under excitation density of $3 \cdot 10^3$ W/cm$^2$ was lower than 600 °C [9], while for $Yb_2O_3$ nanocrystals under excitation density of $2 \cdot 10^2$ W/cm$^2$ it was over 1000 °C [15].

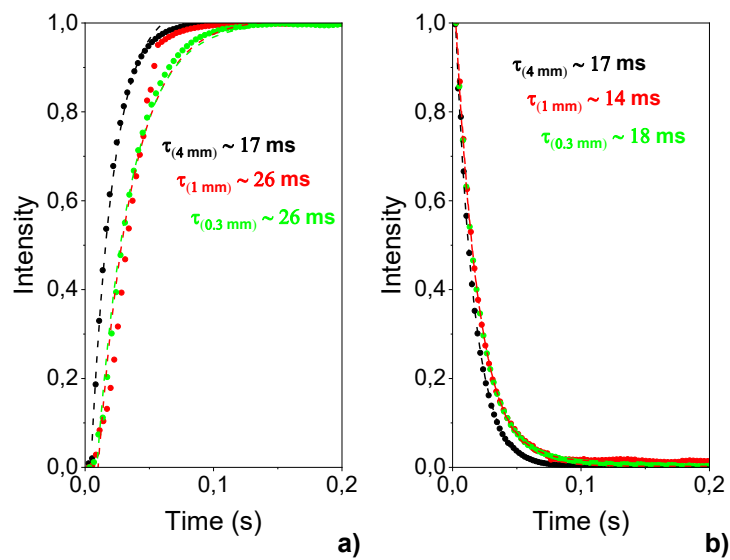

Fig. 2. Temporal evolution of LIWE intensity after switching on (a) and switching off (b) the laser excitation.

One of the characteristics of LIWE process is emission instability in time. Usually the intensity of white emission decreases over time. The origin of this behavior is still unknown. According to the most popular theories it can be caused by saturation of LIWE emission centers or negative influence of the temperature. Most probably, the decrease in the emission intensity is caused by the combination of both these factors.

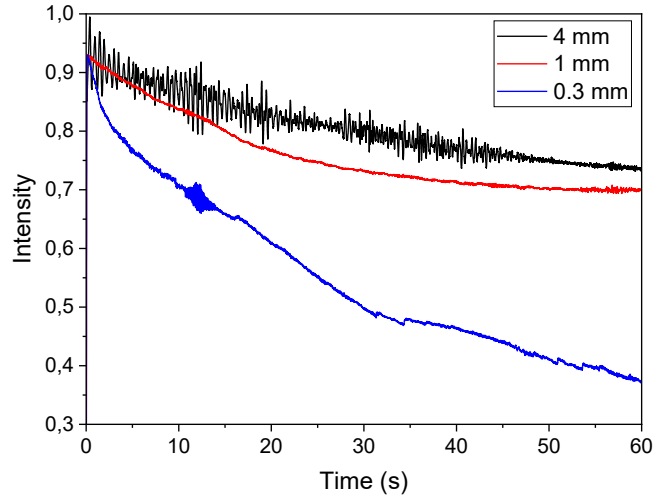

Fig. 3. Temporal evolution of LIWE intensity after switching on the laser excitation obtained for the samples with the thicknesses of 0.3 mm, 1 mm, and 4 mm.

The increase in excitation time of $Cr^{4+}$:YAG transparent ceramics leads to decrease in the LIWE intensity. Beforehand it was proposed that this decrease in the emission intensity is caused by the heating of the samples during LIWE generation [9]. Despite high thermal conductivity of $Cr^{4+}$:YAG transparent ceramics, 85% of energy absorbed by $Cr^{4+}$ ions is converted into heat being a reason of increase of the sample temperature. In order to confirm the negative influence of the temperature on the LIWE intensity, three $Cr^{4+}$:YAG transparent ceramics with thicknesses of 0.3 mm, 1 mm and 4 mm were prepared. The size of the excitation laser spot was 0,17×0.07 mm. LIWE was measured by the irradiation of the sample near its edge, so the ratio of thickness to the laser spot was 2, 6, and 24, respectively. The sample with the lower thickness/spot ratio shows the steepest decline in the intensity emission (**Fig. 3**). This effect can be explained by faster accumulation of heat in samples with lower thickness/spot ratio confirming the negative impact of temperature on LIWE intensity. It should be noted that the LIWE process is unstable, and often the same sample shows the different rates of intensity decrease during different measurements. Also, an increase in the emission intensity was observed in some peculiar cases. However, the general trend remains the same.

## Discussion

A possible cause of LIWE is the recombination of charge carriers during electron transfer in donor/acceptor mixed valence pair. We assume that the electron transfer occurs between chromium ions in different valence states, so-called Intervalence Charge Transfer. The $Cr^{3+}$ and $Cr^{4+}$ ions can act as donor and acceptor, respectively. The transition of the electron into a charge transfer state occurs due to simultaneous absorption of four photons. This energy matches the difference between the $Cr^{4+}$ ground state and the bottom of the conduction band [16]. When this electron is transferred to the Cr ion, charge-transfer luminescence is observed. Changing the valence state of $Cr^{3+}$ to $Cr^{4+}$ ion requires high reorganization energies causing very wide emission bands [17]. The role of the surface in LIWE phenomenon is still unclear. Surface of the YAG is charged due to the difference between the number of oxygen and cation vacancies at the surface [18]. This charge is compensated by the charge of defects of opposite sign from space charge region. Interaction between the space charge region and/or the charged surface with donor/acceptor pair can be a reason of surface nature of LIWE. It should be noted that the exact origin of surface nature of LIWE is still unclear at this stage of the study.

## Acknowledgements

This work was supported by Polish National Science Centre, grant: PRELUDIUM-18 2019/35/N/ST3/01018.

## Conflict of Interest

On behalf of all authors, the corresponding author states that there is no conflict of interest.

## Data availability statement

The datasets generated during and/or analysed during the current study are available from the corresponding author on reasonable request.